\documentstyle[11pt]{article}
\begin{document}
\def\lsim{\, \lower2truept\hbox{${<
\atop\hbox{\raise4truept\hbox{$\sim$}}}$}\,}
\def\gsim{\, \lower2truept\hbox{${>
\atop\hbox{\raise4truept\hbox{$\sim$}}}$}\,}

\begin{center}{\bf COSMOLOGICAL EVOLUTION OF EXTRAGALACTIC SOURCES \

\vspace{0.4truecm}

IN THE INFRARED }

\vspace{0.4truecm}

{\bf AND CONTRIBUTIONS TO THE BACKGROUND RADIATION}

\vspace{1.7truecm}

G. De Zotti$\ ^1$\/, A. Franceschini$\ ^1$\/,
P. Mazzei$\ ^1$\/, G.L. Granato$\ ^2$\/, and L. Danese$\ ^3$\/

\vspace{0.7truecm}
$^1$\it Osservatorio Astronomico,
Vicolo dell'Osservatorio 5, I-35122 Padova, Italy
\vspace{0.5truecm}

\noindent{$^2$\it Dipartimento di Astronomia, Universit\`a di Padova,
Vicolo dell'Osservatorio 5, I-35122 Padova, Italy}

\noindent{$^3$\it SISSA - International School for Advanced Studies,
Strada Costiera, 11, I-34014 Trieste, Italy}
\end{center}

\vspace{2.truecm}

\begin{center} {\bf Abstract}
\end{center}
%\oneskip
%\parindent=1truecm
%\noindent
Infrared surveys provide essential insights on galaxy evolution.
If near-IR studies suggest mild evolution of stellar populations
with cosmic time, indications of a
substantial evolution have been seen in the far-IR,
although the available information is
largely insufficient to delineate precise evolutionary properties.

A consistent picture encompassing all the currently available
data may be obtained assuming that dust extinction hides
the early evolutionary phases of spheroidal galaxies in the optical band,
while the corresponding dust re-radiation in the far-IR may have been,
during the early evolutionary phases, orders of magnitude larger
than today.
Hyperluminous IRAS galaxies might be extreme examples of this situation.
Additional indications that at least some spheroidal galaxies may have been
very dusty during their early evolution are provided by recent data on high
redshift radio galaxies and quasars.

Galaxies are the likely dominant contributors to the IR background.
However, in the framework of unified models for Active Galactic Nuclei, a
large number of nuclei hidden by a dusty torus may be expected. Implications
for the IR background are discussed.

\vfill\eject

%\bigskip\medskip\noindent

\section{\bf  Galaxy evolution in the optical-IR.\
General observational trends}

Optical surveys have constituted for many years the primary tool to
investigate the photometric evolution of galaxies, reflecting the evolution of
stellar populations. CCD detectors have allowed to
reach extremely faint magnitudes, down to $B \simeq 27$--28,
i.e. to
surface densities as high as $\simeq 3\times 10^5$ galaxies deg$^{-2}$.
A remarkable excess over a smooth extrapolation from brighter magnitudes
was observed in the B-band counts, beginning at $B \simeq 20$ and
continuing to the faintest levels.

Counts and colors indicated a substantial increase with look-back time of
the surface density of blue galaxies, consistent with evolutionary stellar
population synthesis models
provided that the universe has a low density ($q_0 \simeq 0.05$) and galaxies
form at high redshifts ($z_{\rm for} \simeq 30$).

The observed redshift distributions of galaxies down to $B \simeq 22$--24,
however, cannot be accounted for by pure luminosity evolution models,
at least if the local luminosity function keeps relatively flat at its faint
end. Various
explanations have been proposed, resorting to either density evolution
\cite{Rocca--VolmerangeGuiderdoni90} or to luminosity dependent
evolution \cite{BroadhurstEllisShanks88}, or to a non-zero
cosmological constant \cite{Fukugitaetal90}; on the other hand,
Gronwall and Koo \cite{GronwallKoo} obtained good fits to the data
with a standard photometric evolution model with a SMC extinction law,
assuming a strong excess (still consistent with the data) of low luminosity
galaxies over the Schechter luminosity function.

Crucial constraints, concerning in particular the effects of dust in high-$z$
galaxies \cite{Franceschinietal94},
come from observations at infrared wavelengths.
A great observational stride occurred thanks to the impressively
fast advances in infrared array technology, which have already made
possible to reach a limiting magnitude $K = 23$
\cite{Gardneretal93,Djorgowskietal1995},
i.e. surface  densities $\simeq 2\times 10^5\,\hbox{deg}^{-2}$
close to those of the deepest optical surveys.

Unlike optical counts, which are highly responsive to young stellar
populations, hence prone to very uncertain K-corrections and biased against
early-type galaxies, near-infrared counts
are much less sensitive to the detailed star-formation history.
By allowing a more straightforward interpretation, they are going to set
much tighter constraints on galaxy evolution.

The extremely interesting new result was
that counts in the K band show a remarkable flattening below $K \simeq 20$,
at variance with the steep counts observed in the optical.

In the far-IR, the first indication that the IRAS galaxy population evolves
significantly came from $60\,\mu$m counts performed at the faintest flux
levels \cite{Hacking and Houck,Ashbyetal} and at medium sensitivities
\cite{Saunders et al}.
The more detailed information made available by redshift surveys
\cite{Oliver et al1995} suggests a strongly evolving luminosity function, with
a luminosity evolution rate comparable to that of quasars:
$$L(z)=L_0(1+z)^{3.3}. \eqno(1)$$
The redshift space, however, was sampled by the IRAS survey only to $z\sim
0.2$, not enough to distinguish between luminosity and density evolution.
The limited sampled volume also implies that the estimate of the
evolution rate in eq. (1) is  possibly affected by large-scale structure. An
underdensity by a factor of roughly two in the local universe has been invoked
to interpret the very steep counts of galaxies in the optical at $B<18$, in
particular from the APM survey in the southern sky \cite{Ellis 1995}.

Near- and far-IR selections emphasize quite different physical processes
in galaxies: photospheric emission of red giant stars, mostly unaffected by
diffuse dust, in the K band, and dust re-radiation
powered by young stellar
populations at far-IR wavelengths. To best exploit the information content of
the two bands, we need a tool to relate the properties of the ISM (gas
fraction, metal abundance and dust content) with those of the stellar
populations. The next chapter is devoted to discuss such a tool.

\section  {\bf A comprehensive view on galaxy evolution: the
basic recipes}

Evolutionary population synthesis (EPS), models which account for dust effects,
have been worked out by \cite{Mazzeietal1992,Mazzeietal1994}.
Given their large spectral
coverage, from ultraviolet to 1 mm, these models
allow a synoptic view of data relevant
to understand galaxy evolution. They deal, in a self-consistent way, with
chemical and photometric evolution of a galaxy over four decades in
wavelength. The synthetic spectral energy distribution (SED) incorporates
stellar emission, internal extinction and re-emission by dust. The stellar
contribution, including all evolutionary phases of stars born with different
metallicities, extends up to $25\,\mu$m. Dust includes a cold component,
heated by the general radiation field,  and a warm component associated with
HII regions.
Emission from policyclic aromatic hydrocarbon molecules (PAH) and
from circumstellar dust shells are also taken into account.

The star formation rate (SFR) and the initial mass function (IMF) are the basic
input functions. The most critical parameter, whose
variation is able to account for the whole Hubble sequence, is the initial
value of the SFR, $\psi_0$. Low values of $\psi_0$ correspond to a SFR slowly
decreasing with time, the standard scenario for late--type systems; high values
of $\psi_0$ provide a drastically different behaviour, due to the fast
decreasing of the gas fraction, hence of the SFR: this corresponds to
early--type systems.

Models successfully match the SED of local galaxies of different morphological
type over four decades in
frequency (\cite{Mazzeietal1992,Mazzeietal1994}).
In the following we will refer to models
computed  with a Salpeter IMF and a SFR depending on a power law of
the mass of gas: SFR $\propto M_{gas}^n$, with $n=0.5$. This corresponds to a
quick metal enrichment and a prompt contamination by dust of the interstellar
medium: early--type systems become bright far--FIR sources given the high
optical depth provided by these models during their initial evolution.

\section{\bf Observational tests of the evolution}

The evolution with cosmic time of stellar populations in galaxies can be
directly investigated by looking at the spectral energy distribution (SED)
of galaxies at different redshifts. ``Normal'' high-$z$ galaxies,
however, proved to be remarkably elusive: none of the 100 (out of a total of
104) galaxies with measured redshift in the sample of Colless et al.
\cite{Collessetal1993},
complete to $B=22.5$, has $z>0.7$; the highest redshift in the complete
$B <24$ sample of Cowie et al. \cite{Cowieetal91} is $z=0.73$; the
measured redshifts of near-IR selected galaxies down to $K=20$
\cite{Songailaetal94} (nearly 100\%
complete at $K<18$ and $\sim 70\%$ complete at $K=19$--20) are $\lsim 1$,
with the exception of one object at $z=2.35$.

So far, the only effective method to find high-$z$ galaxies has been
optical identifications of radio sources \cite{McCarthy93}; far-IR/sub-mm
observations of these objects are providing extremely important indications
on the properties of young galaxies.

In the following, we will briefly review recent observational results
in the near-IR and far-IR/sub-mm bands.

\subsection {\it Near-IR--selected galaxy samples}

Impressive results have been recently obtained
by the IfA team at Hawaii \cite{Songailaetal94},
\cite{Cowieetal94}.
Figures 1 and 2 compare the data of \cite{Songailaetal94} with synthetic
colours yielded by models described in the previous section.

We note that evolutionary effects in the spectra of early--type systems
cannot be neglected at $z\ge 1$; the use of template SED's
of local objects
to infer photometric redshifts of galaxies lacking spectroscopy may be
misleading. We have compared, in particular, the colours of the 16
unidentified galaxies in the K-band selected sample by \cite{Songailaetal94}
and found that most of them are likely to be at $z>1$.
A higher SFR implies bluer colours at $z>1$ than expected by simply
reshifting local SED's.

Note, in any case, that both the red peak in the colours at $z\simeq 1$ and the
subsequent blueing, strongly depend on the adopted cosmological model; in
particular, colours $(I-K) > 5\,$ are difficult to obtain in a closed world
model.

Indications in favour of a merging-driven
evolution, in which galaxies become more numerous but less luminous in the
past, have been found e.g. by \cite{Cowieetal94}.
Alternatively, we find that, taking into account the combined spectral
effects of the evolving stellar populations and of extinction by dust,
all the basic statistical properties of K-band selected galaxies can be
explained by pure luminosity evolution of the observed local luminosity
functions. We account, in particular, for the blueing trend observed at
$K>19$ and $z \sim 1$ as an evolutionary effect due to the higher SFR, rather
than a change in the galaxy population mixture.

A world model with $q_0=0.15$ and $z_{\rm form}=7$
for early--type and $z_{\rm form}\simeq 2$
for late-type galaxies is indicated by fits of the observed counts,
local luminosity functions, redshift and colour distributions.

%\bigskip\noindent
\subsection {\it Far-IR and radio-selected samples}

Early--type models including a dusty phase during their initial evolution also
match  successfully the observed spectrum  of the ultraluminous
galaxy IRAS $F10214+4724$ at z=2.29
\cite{MazzeiDeZotti1994}. In Fig. 3 we compare the observed SED with
model predictions; a non-thermal contribution \cite{GranatoDanese1994}
accounting for $60\%$ of
the rest--frame flux at $\lambda=0.1\,\mu$m, has also been included.

Fig. 4 (panels a--c) compares, in the rest--frame, the data on three distant
radio galaxies having recent sub-mm measurements or upper limits
\cite{Ealesetal1993,Hughes1994} with our synthetic SEDs.
The available data can be fully accounted for by ''opaque''
models like those  used by \cite{MazzeiDeZotti1994} to fit the spectrum
of IRAS $F10214+4724$. The inferred galactic ages are $\le 0.1\,$Gyr,
thus alleviating the problem of galaxy ages uncomfortably
high in comparison with the age of the universe as well as the difficulty
to understand the alignement between optical/infrared continua and the
(presumably short lived) radio jets \cite{EisenhardtChokshi90}.

\section {\bf  IR emission of AGN's }

% 	Simple units \def\kpc{{\rm\thinspace kpc}}
\def\Lsun{\hbox{$\rm\thinspace L_{\odot}$}} \def\Mpc{{\rm\thinspace Mpc}}
\def\Msun{\hbox{$\rm\thinspace M_{\odot}$}} \def\col{\hbox{${\rm cm}^{-2}\,$}}
\def\ergps{\hbox{${\rm erg}\,{\rm s}^{-1}$}} \def\ergpsphz{\hbox{${\rm erg}\,
{\rm s}^{-1}\, {\rm Hz}^{-1}$}} \def\ergpspmpc{\hbox{${\rm erg}\, {\rm
s}^{-1}\, {\rm Mpc}^{-3}$}} \def\ergpcmsqps{\hbox{${\rm erg}\, {\rm cm}^{-2}\,
{\rm s}^{-1}$}} \def\kmpspmpc{\hbox{${\rm km}\, {\rm s}^{-1}\, {\rm
Mpc}^{-1}$}} \def\kmps{\hbox{$\km\s^{-1}\,$}}

The infrared luminosity is often a significant fraction of the bolometric
luminosity of AGN's.
In the case of radio quiet AGN's, the origin of the IR emission
is still debated, although in the last few years observational evidence has
accumulated in favour of  dust reprocessing  of the primary optical--UV
emission. Major issues in favour of dust emission are
(cf. \cite{GranatoDanese1994} and references therein): (i) the observed
SED's showing a steep rise of the submillimetric continuum
between 1000 and 100 $\mu$m which is far too steep for
any current non--thermal model, and a local minimum in $\nu F_{\nu}$ at
$\lambda \simeq 1 \mu$m, which is naturally explained by
models wherein dust radiates with a limiting sublimation temperature $T_s\sim $
1500 K; (ii) the IR variability and its relationship with UV and optical
variability with delays among different bands which are easily interpreted in
dust models; and (iii) the observed low level of polarization.

Presence of optically thick dust with azimuthal symmetry has been also
advocated to explain the observed difference between broad-- and narrow--line
AGN in the framework of unified schemes (see \cite{Antonucci1993}
for a comprehensive
review). As is well known, spectropolarimetric data have shown that broad
lines also are  visible in polarized light in narrow-line nuclei.
More recently, infrared spectroscopy in
the near--IR revealed the presence of broad components in hydrogen IR lines
(see e.g. \cite{GoodrichVeilleuxHill1994}),
suggestive of the possibility of a direct
view of broad--line regions in the IR. Less
direct evidence of dust around active nuclei has been found through studies of
ionization cones in several Seyfert 2 galaxies, which demonstrate that the gas
in the host galaxies sees anisotropic ionizing sources. Also X--ray
observations revealed high absorbing columns ($N_{HI} \sim 10^{23}$--
$10^{25}\hbox{cm}^{-2}$) in many Seyfert 2 galaxies, whereas
Seyfert 1 X-ray spectra exhibit moderate to low absorption.

Structure and  emission of dust tori or warped discs around AGN's have been
investigated by several authors (\cite{PierKrolik1993,GranatoDanese1994}).
In the case of tori, dust is assumed to have standard galactic composition,
size distribution and sublimation temperature. Various structures have been
investigated depending on equatorial optical depth, covering factor, shape,
radial and axial dust density distribution. The models can be broadly divided
in two classes. A first one includes models with extremely high optical depths
($A_V\ge 500$) and extremely compact structure (the ratio of the outer
$r_{\rm out}$ to the inner radius $r_{\rm in}$ of the dust distribution
ranges from few
to several tens). Typical problems of these models are that torus emission does
not fit the observed spectra and surface brightnesses of Seyfert 1 and 2
unless additional dust components are invoked (see e.g. \cite{PierKrolik1993}).

The second family of models requires less extreme optical depth ($A_V\sim 100$)
and more extended tori ($r_{\rm in}/r_{\rm out}\sim $ few hundreds)
\cite{GranatoDanese1994}.
These models have the nice properties of fitting the observed SED's
and of predicting a size of the region emitting at
mid--IR extending over several
tens of parsecs in agreement with the observations.
Moreover the predicted anisotropy of the 10$\mu$m emission is within the limits
required by the comparison of the luminosity functions of type 1 and type 2
Seyferts \cite{GiuricinMardirossianMezzetti1995}.

If the latter models hold, the $12\,\mu$m luminosity functions of
the two Seyfert types
\cite{Rushetal1993} and the statistics of the ratio of hard X--ray to
$12\,\mu$m fluxes \cite{Barconsetal1995} indicate
that not many hidden AGN's exist. At most 20$\%$ of
Seyfert 1 and 50$\%$ of Seyfert 2 galaxies could have been misidentified as
non--active. Moreover the non--linear relation $L_{12}\propto L_{HX}^{0.7}$
suggests that the covering factor decreases with increasing source power
\cite{Barconsetal1995}.

\section {\bf Contributions to the background flux}

In the absence of dust extinction, the diffuse radiation generated by
early stellar nucleosynthesis is expected to peak in the near-IR and to have
a bolometric intensity:
$$I_{\rm bol}={0.007 c^3 \rho f_m \over 4\pi (1+z_F)}\
\hbox{erg}\,\hbox{cm}^{-2}\,\hbox{s}^{-1}\,\hbox{sr}^{-1}, \eqno(2)$$
where $z_f$ is the galaxy formation
redshift, $\rho$ is the density of the processed material to produce a mass
fraction $f_m$ of metals. Dust eventually re-radiates this energy at longer
wavelengths.

Estimates of the IR background intensity have been attempted in the near-IR
and sub-mm bands, where the bright foreground emissions from the Galaxy and
Inter-Planetary Dust (IPD) are at a minimum. In any case, careful modelling
and subtraction of foregrounds is necessary. A summary of recent results
is shown in Fig. 5.

%\bigskip\noindent
\subsection {\it The near-IR background}

The thick solid line labelled ``d'' in Fig. 5 shows the contribution
of starlight as implied by models described in \S3.1. This estimate is
quite a robust one, constrained as it
is by existing deep counts of galaxies in K band.
The predicted spectral shape is also well constrained up
to $\lambda \sim 5 \mu$m by our present understanding of photometric
evolution.

The tightest observational limits come from observations of very high-energy
(TeV) $\gamma$-rays in the spectra of the BL Lac object Mkn~421 and
follow from the fact that
a flux of extragalactic $\gamma$-rays is attenuated by interactions with
ambient photons leading to the production of electron-positron pairs
(\cite{SteckerandDeJager1993}, \cite{DvekandSlavin}). These limits are
consistent with the near-IR background mostly originating from ordinary stellar
processes in galaxies.

Also reported in Figure 5 is the estimated final sensitivity of DIRBE on COBE,
which, in principle, would allow accurate determinations
of the extragalactic background flux from 1 to 200 $\mu$m. As already
mentioned, however, the main limiting factor are
uncertainties in the estimates of the foreground components.

%\bigskip\noindent
\subsection {\it The far--IR/sub--mm background}

In Fig. 5, the lines labelled ``a'', ``b'', ``c'' show predictions of
detailed evolutionary models, while
the thin line is based on simpler kinematical prescriptions fitting the
$60\,\mu$m statistic \cite{Franceschinietal1991}. We see an appreciable
latitude in such predictions. Note, however, that the extreme curves ``a'' and
``c'' bound a sort of conservative ``allowed region'', the former curve,
assuming no
evolution,  being already inconsistent with $60\,\mu$m counts, the latter
corresponding to an evolution rate probably inconsistent with the observed lack
of high-$z$ galaxies among faint IRAS sources\cite{Ashbyetal}.

The prediction of the kinematic model implies an higher flux level at mm
wavelenghts as
it assumes a dust temperature distribution constant with time. Physical models
predict a lower mm background because dust gets warmer in the past, following
the increase of the SFR and of the average radiation field in galaxies.

The various observational limits, set by re-analyses of IRAS survey data
\cite{Oliver et al1995}, by the COBE-FIRAS all-sky spectra deconvolved from
the Galaxy emission \cite{Wrightetal} and by rocket observations
\cite{Kawadaetal1994}, are within a factor of a few of the expected
extragalactic flux. The
latter turns out to be very close to the sensitivity limits of current
observations.

%\bigskip\noindent
\subsection {\it A mid-IR background from hot dust in AGNs?}

At the minimum between the predicted {\it starlight} background in the
near-IR and the {\it dust re-radiation} feature in the far-IR, there is a
spectral region ($5<\lambda<20\,\mu$m) in which properties of even local
galaxies are very poorly known (essentially because of the poor sensitivity
of IRAS at these wavelengths). Existing data suggest, in any case, that the
integrated emission of normal galaxies should keep below $3\times 10^{-6}\,
\hbox{erg}\,\hbox{cm}^{-2}\,\hbox{s}^{-1}\,\hbox{sr}^{-1}$,
unless dust with implausibly high temperature
characterizes the early evolution phases.

As previously
discussed, dust hot enough to emit at these wavelengths is likely to be
present in AGNs. The corresponding
contribution to the mid-IR background is set by the convolution of the local
IR volume emissivity of AGNs with their evolution rate. We report in
Fig.~5 (line marked AGN) a crude estimate based on the $12\,\mu$m
luminosity function of type 1 and 2 Seyfert galaxies by \cite{Rushetal1993}
and an
evolution rate equal to that of optical QSOs (eq.[1]), extrapolated to
$z_{\rm max}=4$ with $q_0=0.05$. The adopted IR spectra for the two AGN types
are the average SED of Seyfert 1 nuclei and the SED of NGC 1068
\cite{GranatoDanese1994}. The IR luminosity function
of Seyferts 2 has been renormalized upwards by a factor 4
to account for the fraction of optically undetected, dust extinguished AGNs
predicted by the unified AGN model. We see that, with these somewhat extreme
ingredients, the contribution of dusty AGNs in the range 10--50$\,\mu$m may
exceed those of other galaxy populations.

%
%     REFERENCES
%
%%%%%%% reference definitions
%
\def\aa #1 #2{{\it Astr. Astrophys.,}~{\bf #1}, {#2}}
\def\aar #1 #2{{\it Astr. Astrophys. Rev.,}~{\bf #1}, {#2}}
\def\aas #1 #2{{\it Astr. Astrophys. Suppl.,}~{\bf #1}, {#2}}
\def\araa #1 #2{{\it Ann. Rev. Astr. Astrophys.,}~{\bf #1}, {#2}}
\def\aj #1 #2{{\it Astr. J.,}~{\bf #1}, {#2}}
\def\alett #1 #2{{\it Astrophys. Lett.,}~{\bf #1}, {#2}}
\def\apj #1 #2{{\it Astrophys. J.,}~{\bf #1}, {#2}}
\def\apjs #1 #2{{\it Astrophys. J. Suppl.,}~{\bf #1}, {#2}}
\def\ass #1 #2{{\it Astrophys. Space Sci.,}~{\bf #1}, {#2}}
\def\baas #1 #2{{\it Bull. Am. astr. Soc.,}~{\bf #1}, {#2}}
\def\ca #1 #2{{\it Comm.Astr.,}~{\bf #1}, #2}
\def\fcp #1 #2{{\it Fundam. Cosmic Phys.,}~{\bf #1}, {#2}}
\def\memsait #1 #2{{\it Memorie Soc. astr. ital.,}~{\bf #1}, {#2}}
\def\mnras #1 #2{{\it Mon. Not. R. astr. Soc.,}~{\bf #1}, {#2}}
\def\qjras #1 #2{{\it Q. Jl R. astr. Soc.,}~{\bf #1}, {#2}}
\def\nat #1 #2{{\it Nature,}~{\bf #1}, {#2}}
\def\pasj #1 #2{{\it Publs astr. Soc. Japan,}~{\bf #1}, {#2}}
\def\pasp #1 #2{{\it Publs astr. Soc. Pacif.,}~{\bf #1}, {#2}}
\def\physl #1 #2{{\it Phys.Lett.,}~{\bf #1}, #2}
\def\physrep #1 #2{{\it Phys.Rep.,}~{\bf #1}, #2}
\def\physreva #1 #2{{\it Phys. Rev. A,}~{\bf #1}, {#2}}
\def\physrevb #1 #2{{\it Phys. Rev. B,}~{\bf #1}, {#2}}
\def\physrevd #1 #2{{\it Phys. Rev. D,}~{\bf #1}, {#2}}
\def\physrevl #1 #2{{\it Phys. Rev. Lett.,}~{\bf #1}, {#2}}
\def\pl #1 #2{{\it Phys. Lett.,}~{\bf #1}, {#2}}
\def\prsl #1 #2{{\it Proc. R. Soc. London Ser. A,}~{\bf #1}, {#2}}
\def\ptp #1 #2{{\it Prog. theor. Phys.,}~{\bf #1}, {#2}}
\def\ptps #1 #2{{\it Prog. theor. Phys. Suppl.,}~{\bf #1}, {#2}}
\def\rmp #1 #2{{\it Rev. Mod. Phys.,}~{\bf #1}, {#2}}
\def\sovastr #1 #2{{\it Sov. Astr.,}~{\bf #1}, {#2}}
\def\sovastrl #1 #2{{\it Sov.Astr. (Lett.),}~{\bf #1}, L#2}
\def\ssr #1 #2{{\it Space Sci. Rev.,}~{\bf #1}, {#2}}
\def\va #1 #2{{\it Vistas in Astronomy,}~{\bf #1}, {#2}}
\def\book #1 {{\it ``{#1}'',\ }}
%%%%%%%%%%%%%%%%%%%%%%%%%%%%%%%%%%%%%%%%%%%%%%%%%%%%%%%%%%%%%%%%%%%%

\vfill\eject

\centerline{\bf Figure captions}

\begin{description}

\item{\bf Fig. 1} Colour-redshift relationships predicted by photometric
evolution models described in \S2 for galaxies of different morphological
types assumed to start their star formation at the redshifts $z_f$ indicated,
compared with data of \cite{Songailaetal94}. The open symbols show the colours
of galaxies lacking redshift measurements, plotted at a nominal $z=5$.

\item{\bf Fig. 2} Colour-colour plots for intermediate redshift ($0.5\leq z
\leq 1$) galaxies (filled circles) and of candidate high-$z$ galaxies (open
circles) in the sample of \cite{Songailaetal94}. The lines show
predictions of the same models as in Fig. 1. The circle and the star mark
the synthetic colours at $z=1$ and $z=2$, respectively, of galaxies of
each morphological type.

\item{\bf Fig. 3} Spectral energy distribution of the hyperluminous
galaxy IRAS $F10214+4724$ at $z=2.29$ (see \cite{MazzeiDeZotti1994}
for details and references). A non-thermal contribution having a spectrum
equal to the mean spectrum of Seyfert 1 nuclei \cite{GranatoDanese1994} and
accounting for $60\%$ of
the rest--frame flux at $\lambda=0.1\,\mu$m, has been included.

\item{\bf Fig. 4} Fits to the SED's of three high-$z$ radiogalaxies. The
ages of the models are specified in each panels. The lower thin
curves show the assumed nuclear contributions.

\item{\bf Fig. 5} Extragalactic background light at IR to sub-mm wavelengths.
Curve $a$ corresponds to no evolution, curves $b$
and $c$ to moderate or strong extinction, respectively, of early type galaxies
during early evolutionary phases; curve $d$ is the integrated starlight of
distant galaxies in the near-IR, derived from models fitting the deep
K-band counts. Curve $S$ shows the contribution of
stars in our own Galaxy, at high galactic latitudes. The curve
labelled ``AGN'' shows the somewhat extreme estimate of the possible
AGN contribution to the background described in \S5.2.
The DIRBE sensitivity as function of $\lambda$ is also shown.

\end{description}
\end{document}